\documentclass[aps,prb,twocolumn,amsmath,amssymb,superscriptaddress,scrartcl,eqsecnum,longbibliography]{revtex4-1}
\usepackage{extarrows}
\usepackage{slashed}
\usepackage{amsmath,amsfonts,amssymb,graphics,graphicx,epsfig,color,times,indentfirst,layout}
\usepackage{lipsum}
\usepackage{mathrsfs}
\usepackage[unicode=true,pdfusetitle,bookmarks=false,colorlinks=true,citecolor=black,urlcolor=black,linkcolor=black]{hyperref}
\usepackage{tikz}
\usepackage{tikz-cd}
\usetikzlibrary{arrows}
\usetikzlibrary{intersections}
\usetikzlibrary{shapes.geometric}

\def\be{\begin{equation}}
\def\ee{\end{equation}}

\begin{document}

\rightline{MIT-CTP/4983}

\title{
Quantum dynamics in sine-square deformed conformal field theory:\\
Quench from uniform to non-uniform CFTs
}

\date{\today}

\author{Xueda Wen}

\affiliation{Department of Physics, Massachusetts Institute of Technology, Cambridge, MA 02139, USA}

\author{Jie-Qiang Wu}
\affiliation{Center for Theoretical Physics, Massachusetts Institute of Technology, Cambridge MA, 02138 USA}

\begin{abstract}
In this work, motivated by the sine-square deformation (SSD) for (1+1)-dimensional quantum critical systems,
we study the non-equilibrium quantum dynamics of a conformal field theory (CFT) with SSD,
which was recently proposed to have continuous energy spectrum and continuous Virasoro algebra.
In particular, we study the time evolution of entanglement entropy after a quantum quench
from a uniform CFT, which is defined on a finite space of length $L$, to a sine-square deformed CFT.
We find there is a crossover time $t^{\ast}$ that divides the entanglement evolution into two interesting regions.
For $t\ll t^{\ast}$, the entanglement entropy does not evolve in time;
for $t\gg t^{\ast}$, the entanglement entropy grows as $S_A(t)\simeq \frac{c}{3}\log t$,
which is independent of the lengths of the subsystem and the total system.
This $\log t$ growth with no revival indicates that a sine-square deformed CFT effectively has an infinite length,
in agreement with previous studies based on the energy spectrum analysis.
Furthermore, we study the quench dynamics for a CFT with M$\ddot{\text{o}}$bius deformation,
which interpolates between a uniform CFT and a sine-square deformed CFT.
The entanglement entropy oscillates in time with period $L_{\text{eff}}=L\cosh(2\theta)$,
with $\theta=0$ corresponding to the uniform case and $\theta\to \infty$ corresponding to the SSD limit.
Our field theory calculation is confirmed by a numerical study on a (1+1)-d critical fermion chain.
\end{abstract}
\maketitle


\section{Introduction}

(1+1)-dimensional quantum many-body systems with sine-square deformation (SSD)
 have been extensively studied in recent years. \cite{Gendiar2009,Gendiar2010,Hikihara2011,Gendiar2011,
Shibata2011,Shibata2012,Shibata2013,Katsura2011a,
Katsura1110,Maruyama2011,
Tada1404,Okunishi2015,Tada1504,Tada1602,Okunishi1603,Ryu1604,Katsura1709,Tada1712}
The SSD was originally introduced as a spatial deformation of Hamiltonian density that efficiently
suppresses the boundary effects.\cite{Gendiar2009,Gendiar2010,Hikihara2011,Gendiar2011}
The set-up is as follows:
Consider a (1+1)-d quantum many-body system with open boundary condition and Hamiltonian
\be\label{H00}
H_0=\int_0^L h(x)dx,
\ee
where $h(x)$ is the Hamiltonian density. For simplicity, we assume $h(x)$ is uniform.
Now we deform the Hamiltonian as follows
\be\label{SSD}
H_{\text{SSD}}=\int_0^L 2\sin^2\left(\frac{\pi x}{L}\right) h(x) dx.
\ee
Apparently, the system is disconnected at $x=0$ $(L)$.
This kind of deformation shows remarkable properties for (1+1)-d quantum critical systems.
It was found that the ground state of (\ref{SSD}) is identical to that of a uniform system with periodic boundary condition
within numerical accuracy.\cite{Hikihara2011,Gendiar2011,Shibata2011,Shibata2012,Shibata2013}
This property was further verified analytically in some exactly solvable models.
\cite{Katsura2011a,Katsura1110,Maruyama2011,Okunishi2015}

Later, the SSD of a two dimensional conformal field theory (CFT) was investigated in Ref.\onlinecite{Katsura2011a},
where it is found that the Hamiltonian for a generic CFT with SSD can be expressed as
$\mathcal{L}_0+\bar{\mathcal{L}}_0$, where
\be\label{SSDf}
\mathcal{L}_0=L_0-\frac{L_1+L_{-1}}{2},
\ee
and similarly for $\bar{\mathcal{L}}_0$.
Here $L_n$ ($n=0,\pm 1, \cdots$) are Virasoro generators in a CFT.
\footnote
{See Eqs. (\ref{H0}) and (\ref{Hpm}) for an explicit definition of the Hamiltonian in terms of stress-energy tensors.
}
Considering the Hamiltonian (\ref{SSD}), since $\sin^2(\frac{\pi x}{L})$ vanishes at the boundary,
there is no difference in
 the system between an open boundary condition and a periodic boundary condition. On the other hand, for periodic boundary condition the Hamiltonian (\ref{SSDf}) has the same ground state as a uniform CFT with Hamiltonian $L_0+\bar{L}_0$,
because in periodic system the CFT vacuum is annihilated by $(L_1+L_{-1})/2$
and $(\bar{L}_1+\bar{L}_{-1})/2$. With these two observations, the ground state of the SSD system with open boundary condition is the same as that for a uniform system with periodic boundary condition.

In two recent papers by Ishibashi and Tada,\cite{Tada1504,Tada1602}
they studied the sine-square deformed CFT by dipolar quantization.
Choosing different time slices and time translations, they showed that 2d CFTs with SSD have different quantization
other than the radial quantization, which is called dipolar quantization.
They also found a continuous Virasoro algebra which is labeled by a continuous real index $\kappa$:
\be\label{Virasoro_SSD}
[\mathcal{L}_{\kappa},\mathcal{L}'_{\kappa}]=(\kappa-\kappa')\mathcal{L}_{\kappa+\kappa'}+\frac{c}{12}\kappa^3\delta(\kappa+\kappa'),
\ee
with $\kappa>0$, and here $c$ is the central charge of the CFT.
This continuous Virasoro algebra results in a continuous spectrum in the CFT with SSD,
which implies that the sine-square deformed CFT effectively has an infinite length,
even though it is defined on a finite space.

To further understand the property of CFTs with SSD, some regularization schemes were proposed
recently. \cite{Okunishi1603,Ryu1604}
It was found that the connection between the uniform system and the SSD system can be built
by considering the parameterized Hamiltonian
$\mathcal{L}_0+\bar{\mathcal{L}}_0$ with\cite{Okunishi1603}
\be\label{Mobius_L0}
\mathcal{L}_0= L_0-\tanh(2\theta)\frac{L_1+L_{-1}}{2},
\ee
and similarly for $\bar{\mathcal{L}}_0$.
Here one can choose $\theta\ge 0$ without loss of generality.
It can be checked that the rescaled Hamiltonian $\cosh(2\theta)\cdot\left(\mathcal{L}_0+\bar{\mathcal{L}}_0\right)$
is related to the conventional one $L_0+\bar{L}_0$ by a
M$\ddot{\text{o}}$bius transformation.\cite{Okunishi1603}
For this reason, we will refer to the Hamiltonian defined through Eq.\eqref{Mobius_L0} as the M$\ddot{\text{o}}$bius Hamiltonian,
and we will call this kind of deformation as M$\ddot{\text{o}}$bius deformation.
It is noted that for M$\ddot{\text{o}}$bius deformation, one has a M$\ddot{\text{o}}$bius quantization that
bridges the radial quantization for the conventional case and the dipolar quantization for the SSD case.
One can refer to Ref.\onlinecite{Okunishi1603} for more details.
Then $\theta=0$ corresponds to the uniform case, and $\theta\to \infty$ corresponds
to the SSD limit. Within M$\ddot{\text{o}}$bius deformation, one can find a Virasoro algebra which is the same as
Eq.\eqref{Virasoro_SSD}, except that now $\kappa$ is not a continuous real index but it has the expression
$
\kappa=\frac{n}{\cosh(2\theta)},
$
where $n$ is an integer.
In the SSD limit $\theta\to \infty$, $\kappa$ becomes continuous, which results in a continuous spectrum
as observed by Ishibashi and Tada.\cite{Tada1504,Tada1602}

\subsection{Our motivation}

Based on the introduction above, now let us state our motivations in this work:

\begin{enumerate}

\item
Given the continuous energy spectrum for a CFT with SSD, how does it affect the non-equilibrium quantum dynamics?
To avoid any operator dependence, we will study the entanglement evolution in quench dynamics.
It is interesting to see if there is any universal feature in the entanglement evolution, and if yes, how it
is related with the continuous spectrum. As far as we know, there is no work
studying the entanglement property of a CFT with SSD in the non-equilibrium case.

\item

Since the  M$\ddot{\text{o}}$bius Hamiltonian [see Eq.\eqref{Mobius_L0}]
interpolates between the uniform system and SSD system. It is also interesting to study the
quench dynamics governed by this Hamiltonian, and see how it behaves as we
approach the SSD limit.

\item
There is much recent interest in studying the entanglement property
in CFT in curved spacetime (See, e.g., Refs.\onlinecite{RefaelMoore2004,Javier2016,Dubai1606,Viti1507,Eisler2017,
Dubai2017b,Tonni1712,ramirez2015entanglement,rodriguez2017more}.)
One interesting setup is based on the inhomogeneous Hamiltonian density.\cite{Dubai2017b}
We hope that our work will provide a nice setup and approach in this direction.

\end{enumerate}

The rest of this paper is organized as follows:
In Sec.\ref{QuenchToSSD}, we introduce our setup of quantum quench, and
study the entanglement evolution after a quantum quench from a uniform system
to a SSD system. Then in Sec.\ref{QuenchToMobius}, we study the quench
dynamics for the M$\ddot{\text{o}}$bius case, and see how it connects the
uniform and SSD cases. In Sec.\ref{Conclusion}, we present some discussions and
conclude our work. In Appendix \ref{Appendix_Lattice}, we introduce the lattice
model based on which we do numerical calculations.
In Appendix \ref{Appendix_curve},
we interpret the SSD and M$\ddot{\text{o}}$bius Hamiltonian as a CFT in curved spacetime.

\begin{figure}[tp]
\centering
\includegraphics[width=2.75in]{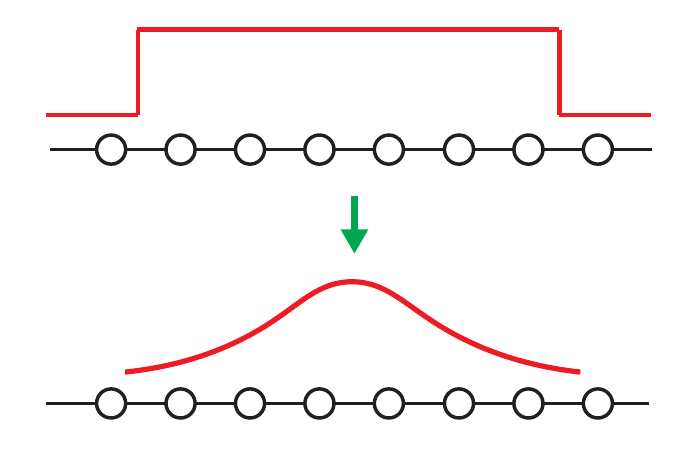}
\caption{
For $t<0$, we prepare our initial state as the ground state of a uniform CFT with open boundary condition.
From $t=0$, we evolve the initial state with a sine-square deformed Hamiltonian.
The red solid lines represent the strength of Hamiltonian density.
The lattice represents a microscopic model that may realize the CFT in IR limit.
}
\label{Scheme}
\end{figure}

\section{Entanglement evolution:\\
Quench from uniform to SSD systems}
\label{QuenchToSSD}

\subsection{Setup}

There are several interesting setups for quantum quenches, such as the
global quench,\cite{calabrese2005evolution,CC_Global,CC2016} local quench,\cite{CC_Local,Nozaki2014,He2014,Stephan2011} and inhomogeneous quantum quench.
\footnote{For a more complete review of recent progress on various quantum quenches in CFTs, one can refer to Ref.\onlinecite{CC2016} and the references therein.}
\cite{Cardy2008,Viti1507,Dubai1606,Eisler2017,BertiniPRL2016,Dubai2017b,BertiniPRL2018,wen2018entanglement,Dubail1712}

Here, we consider a quantum quench from a uniform CFT to a non-uniform CFT.
As shown in Fig.\ref{Scheme}, we prepare our initial state as the ground state $|G\rangle$ of a uniform CFT on
a finite space $[0,L]$, with an open boundary condition.
(The reason we do not choose a periodic boundary condition (PBC) is that, as mentioned in the introduction,
a critical system with a PBC shares the same ground state with the SSD system,
and no quench is expected to happen in this case. It is noted, however, that there may be some
difference in the ground states at UV scale. We are not interested in quantum quench in this case,
the feature of which is non-universal.)
Then at $t=0$, we switch the Hamiltonian
to the sine-square deformed one. Then the time dependent state
can be written as $|\psi(t)\rangle=e^{-iH_{\text{SSD}}t}|G\rangle$.
The correlation function of $O(x_1)\cdots O(x_n)$ at time $t$ can be expressed as
$\langle G|e^{iH_{\text{SSD}}t}O(x_1)\cdots O(x_n)e^{-iH_{\text{SSD}}t}|G\rangle$.

\begin{figure}[tp]
\centering
\includegraphics[width=3.20in]{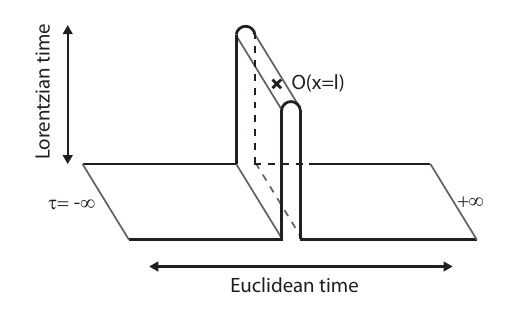}
\caption{
Path integral representation for the correlation function $\langle G | e^{iH_{\text{SSD}}t} \mathcal{O}(x=l) e^{-iH_{\text{SSD}}t} |G\rangle$. The ground state $|G\rangle$ of Hamiltonian $H_0$ can be considered as the path integral
starting from $\tau=-\infty$ and evolving to $\tau=0$ according to Hamiltonian $H_0$.
The width of this strip is $L$.
Then the ground state evolves in the Lorentzian time direction
according to the new Hamiltonian $H_{\text{SSD}}$. After inserting an operator $\mathcal{O}(x)$ at $x=l$, we evolve the state
back to $\tau=0$. Then the state evolves to $\tau=+\infty$ with $H_0$.
}
\label{PI}
\end{figure}

Throughout this work, to study the quench dynamics of a sine-square deformed CFT,
we are interested in the time evolution of entanglement entropy for a subsystem
$A=[0,l]$. The entanglement measure we use is the so-called Renyi entropy
\be\label{RenyiEE}
S_A^{(n)}(t)=\frac{1}{1-n}\log \text{tr}\left[\rho_A^n(t)\right],
\ee
and the von-Neumann entropy
\be
S_A(t)=\lim_{n\to 1}S_A^{(n)}(t).
\ee
The term $\text{tr}(\rho_A^n)$ in $S_A^{(n)}(t)$ is related with the single-point correlation function
of twist operator as follows:\cite{CC2009}
\be\label{Tn}
\text{tr}(\rho_A^n)=\langle G | e^{iH_{\text{SSD}}t} \mathcal{T}_n(x=l) e^{-iH_{\text{SSD}}t} |G\rangle,
\ee
where $\mathcal{T}_n$ is a primary operator with conformal dimension
\be\label{ConformalDimension}
h=\frac{c}{24}\left(n-\frac{1}{n}\right),
\ee
In the following, we will evaluate the correlation function in Eq.\eqref{Tn} with path integral method.
Pictorially, $\langle \psi(t)|\mathcal{T}_n(x=l)|\psi(t)\rangle$ is shown in Fig.\ref{PI} by setting $O(x)=\mathcal{T}_n(x)$.
Note that there are both Euclidean time and Lorentzian time in the path integral.
As shown in the following part, we will do calculation in the Euclidean spacetime by setting $it=\tau$, and
analytically continue back to Lorentzian time in the final step.

\subsection{Entanglement evolution
after a quantum quench from uniform to SSD systems}

To study the entanglement entropy evolution, we will go to Euclidean spacetime by setting $it=\tau$.
Then the correlation function in Eq.\eqref{Tn} has the form
\be\label{Tn_Eu}
\langle G \mid e^{H_{\text{SSD}}\tau} \mathcal{T}_n^{(w)}(w_0,\bar{w}_0) e^{-H_{\text{SSD}}\tau} \mid G \rangle.
\ee
Here in the Euclidean plane $w=\tau+i\sigma$, one has $-\infty<\tau<\infty$, $0\le \sigma\le L$, and
$w_0=0+il$ ($0\le l\le L$) is used to label the position of the twist operator.
The superscript $(w)$ in $\mathcal{T}^{(w)}_n$ denotes the coordinate.
Two conformal boundary conditions are imposed along $\sigma=0$ and $L$, respectively.
For simplicity, we assume the two boundary conditions are the same.
\footnote{If the two boundary conditions along $x=0$ and $x=L$ are different, one
needs to consider the  boundary condition changing operator in the following discussions.}

To evaluate the correlation function in Eq.\eqref{Tn_Eu}, we take the following
two strategies:

(i) Heisenberg picture. Instead of evolving the states,
we will evolve the operator $\mathcal{T}_n$ with the Hamiltonian $H_{\text{SSD}}$ in Heisenberg picture.
By conformal transformation into a certain coordinate, it is quite straightforward to write down the operator's evolution.

(ii) We start from M$\ddot{\text{o}}$bius Hamiltonian $H_{\text{ M$\ddot{\text{o}}$b}}(\theta)$ first. Taking $\theta\to \infty$, we can read out the SSD result. As we mentioned in the introduction, the SSD Hamiltonian has a continuous spectrum. In case of unnecessary IR problem, we regard the M$\ddot{\text{o}}$buis Hamiltonian as a regularization of SSD.
To be concrete, in terms of stress-energy tensor $T$,
the Hamiltonian $H_{\text{ M$\ddot{\text{o}}$b}}(\theta)$ in $w$-plane can be written as\cite{Okunishi1603}
\be\label{MobiusHamiltonian}
H_{\text{M$\ddot{\text{o}}$b}}=
 H_0-\frac{\tanh(2\theta)}{2}\left(
H_++H_-\right),
\ee
where
\be\label{H0}
H_0=\int_0^L \frac{d\sigma}{2\pi}T_{\tau\tau}(\sigma)
=\int_{0}^L\frac{d\sigma}{2\pi}(T(w)+\bar{T}(\bar{w})),
\ee
and
\be\label{Hpm}
H_{\pm}=\int_0^L\frac{d\sigma}{2\pi}\left(
e^{\pm 2\pi w/L}T(w)+e^{\mp 2\pi \bar{w}/L}\bar{T}(\bar{w})
\right).
\ee
Apparently, for $\theta\to \infty$ we have $H_{\text{M$\ddot{\text{o}}$bius}}(\theta\to \infty)=
H_0-\frac{1}{2}(H_++H_-)$, which is the SSD Hamiltonian.\cite{Katsura2011a}

Based on the above two strategies, we are ready to calculate the correlation function in Eq.\eqref{Tn_Eu}.
Readers who are not interested in the technical part can go to the result in Eq.\eqref{SAt_SSD} directly.
Now let us consider the conformal mapping
\be
z=e^{\frac{2\pi}{L}w},
\ee
which maps the strip in $w$-plane to a complex $z$-plane.
The two boundaries along $\sigma=0,\,L$ in $w$-plane are mapped to
a slit along $z=x\pm i0$, with $x\in [0,\infty)$.
The holomorphic part of $H_{\text{M$\ddot{\text{o}}$b}}$ in $z$-plane is
\be
\begin{split}
H_{\text{M$\ddot{\text{o}}$b}}^{(z)}=&\frac{2\pi}{L\cosh(2\theta)}\Big[
\cosh(2\theta)\oint \frac{zdz}{2\pi i}zT(z)\\
&-\frac{\sinh(2\theta)}{2}\oint\frac{z^2+1}{2\pi i}T(z)dz\Big]-\frac{\pi c}{12 L}.
\end{split}
\ee
Apparently, the Hamiltonian $H_{\text{M$\ddot{\text{o}}$b}}^{(z)}$ is still complicate and
we do not know how to act it on the primary field.
It is found that one can use a M$\ddot{\text{o}}$bius transformation
to further map it to $\tilde{z}$-plane:
\cite{Okunishi1603}
\be\label{Mtransform001}
\tilde{z}=f(z)=-\frac{\cosh \theta z-\sinh \theta}{\sinh \theta z-\cosh\theta}.
\ee
Then the holomorphic part of $H_{\text{M$\ddot{\text{o}}$b}}$ has the simple form
\be
\begin{split}
H_{\text{M$\ddot{\text{o}}$b}}^{(\tilde{z})}=&
\frac{1}{iL\cosh(2\theta)}\oint \tilde{z}T(\tilde{z})d\tilde{z}-\frac{\pi c}{12L}\\
=&\frac{2\pi}{L\cosh(2\theta)} L_0^{(\tilde{z})}-\frac{\pi c}{12L}.
\end{split}
\ee
It is similar for the anti-holomorphic part of $H_{\text{M$\ddot{\text{o}}$b}}^{(\tilde{z})}$.
Then we have
\be
e^{H_{\text{M$\ddot{\text{o}}$b}}\tau} \mathcal{T}_n^{(\tilde{z})}(\tilde{z},\bar{\tilde{z}})
e^{-H_{\text{M$\ddot{\text{o}}$b}}\tau}
=
 \lambda^{2h}\mathcal{T}_n^{(\tilde{z})}(\lambda \tilde{z},\lambda\bar{\tilde{z}}),
\ee
which is nothing but the dilatation operation in $\tilde{z}$-plane. Here $h$ is the conformal dimension
of $\mathcal{T}_n$ in Eq.\eqref{ConformalDimension}, and
\be
\lambda:=\exp\left(\frac{2\pi \tau}{L\cosh(2\theta)}\right).
\ee
Back in $z$-plane, its effect is to shift the operator $\mathcal{T}_n^{(z)}$ from $z$ to $z_{\text{new}}$, where
$z_{\text{new}}$ is related with $z$ as
\be
f(z_{\text{new}})=\lambda f(z),
\ee
with $f(z)$ given in Eq.\eqref{Mtransform001}. Then one can obtain
\be
z_{\text{new}}=\frac{[(1-\lambda)\cosh2\theta-(\lambda+1)]z+(\lambda-1)\sinh2\theta}
{(1-\lambda)\sinh2\theta\cdot z+[(\lambda-1)\cosh2\theta-(\lambda+1)]}.
\ee
Therefore, the correlation function of $\mathcal{T}_n$  can be written as:
\begin{small}
\be\label{Tn_Eu_Mob}
\begin{split}
&\langle G \mid e^{H_{\text{M$\ddot{\text{o}}$b}}\tau} \mathcal{T}_n^{(w)}(w_0,\bar{w}_0)
e^{-H_{\text{M$\ddot{\text{o}}$b}}\tau} \mid G \rangle\\
=&
\left(\frac{\partial z}{\partial w}\right)^h
\left(\frac{\partial \bar{z}}{\partial \bar{w}}\right)^h
\left(\frac{\partial z_{\text{new}}}{\partial z}\right)^h
\left(\frac{\partial \bar{z}_{\text{new}}}{\partial \bar{z}}\right)^h
\left\langle
\mathcal{T}_n^{(z)}(z_{\text{new}},\bar{z}_{\text{new}})
\right\rangle.
\end{split}
\ee
\end{small}
Here $\langle \mathcal{T}^{(z)}_n(z,\bar{z})\rangle$ is the one point correlation function in a boundary CFT.
The boundary condition is imposed along the slit $z=x\pm i0$ on real axis, with $x\in [0,\infty)$.
Explicitly,
$\langle \mathcal{T}^{(z)}_n(z,\bar{z})\rangle$ can be expressed as
\be
 \langle \mathcal{T}^{(z)}_n(z,\bar{z})\rangle
=A_n^{\text{b}}\cdot (\frac{1}{4}z^{-\frac{1}{2}}\bar{z}^{-\frac{1}{2}})^h \cdot
(\frac{2a i}{z^{\frac{1}{2}}-\bar{z}^{\frac{1}{2}}})^{2h},
\ee
where $A^{\text{b}}_n$ is an amplitude depending on the selected boundary condition,
which will affect the entanglement entropy by an order $\sim\mathcal{O}(1)$ term.
$a$ is a UV cut-off, which may be considered as the lattice spacing in a microscopic lattice model.

Recall that in the procedures above, the Hamiltonian we use is $H_{\text{M$\ddot{\text{o}}$b}}(\theta)$.
One needs to further take $\theta\to \infty$  to obtain the SSD limit.
After some tedious but straightforward steps, finally we arrive at
\be
\begin{split}
&\langle G \mid e^{H_{\text{SSD}}\tau} \mathcal{T}_n^{(w)}(w_0,\bar{w}_0) e^{-H_{\text{SSD}}\tau} \mid G \rangle\\
=&A^{\text{b}}_n \left(
\frac{2\pi^2 a^2}{L^2}
\right)^h\cdot
\left(
\frac{1}{m(t)^2+m(t)\cdot n(t)}
\right)^h,
\end{split}
\ee
where
\be
m(t)=\sqrt{
n(t)^2+\sin^2\frac{2\pi l}{L}
},
\ee
and
\be
n(t)=\left(1-\cos \frac{2\pi l}{L}\right)\frac{2\pi^2 t^2}{L^2}-\cos\frac{2\pi l}{L}.
\ee
Here we have already taken the analytical continuation $\tau\to it$. Then based on Eqs.\eqref{RenyiEE}$\sim$\eqref{Tn},
one can obtain the entanglement entropy for $A=[0,l]$ as follows
\be\label{SAt_SSD}
S_A(t)=\frac{c}{12}\log \left\{\frac{L^2}{2\pi^2a^2 }\cdot \left[ m(t)^2+m(t)\cdot n(t)\right]\right\},
\ee
where we have neglected the $\mathcal{O}(1)$ term $\sim c\log A_{n=1}^b$ contributed by the conformal boundary condition.
As shown in Fig.\ref{SSD_quench_fig}, we compare our field theory result $S_A(t)$ with the numerical calculation
on a lattice fermion chain (See the appendix for numerics.
The only fitting parameter we used is the global constant shift
in the ground-state entanglement entropy $S_A(t=0)$.
It is noted that for a free fermion model, this
constant term in $S_A(t=0)$ has been exactly evaluated in Ref.\onlinecite{fagotti2011universal}.). They agree in an excellent way.
One remarkable feature is that the entanglement entropy $S_A(t)$ grows as $S_A(t)\simeq \frac{c}{3}\log t$
in the long time limit. Although the system is defined on a finite space $[0,L]$, in contrast to the uniform CFT,\cite{Cardy_revival}
no revival appears here.
This agrees with previous observations that a CFT with SSD effectively has an infinite length limit.
In our case, the signal caused by a quench can never reach the `boundary' of the SSD system and reflect back.
One intuitive picture is to consider the sine-square deformation directly. Since $\sin^2\frac{\pi x}{L}$
vanishes near the boundaries $x=0,\,L$, the local group velocity of quasi-particles
$v(x)=2\sin^2\frac{\pi x}{L}$
will go to zero when
approaching the boundary. Then it takes an infinite time for the quasiparticles to reach the boundary and reflect back.
We will give further discussion on this effect in next section.

It is noted that there is much more information in $S_A(t)$ in Eq.\eqref{SAt_SSD} and Fig.\ref{SSD_quench_fig},
which we will analyze case by case in the following:

\begin{figure}[tp]
\centering
\includegraphics[width=3.5in]{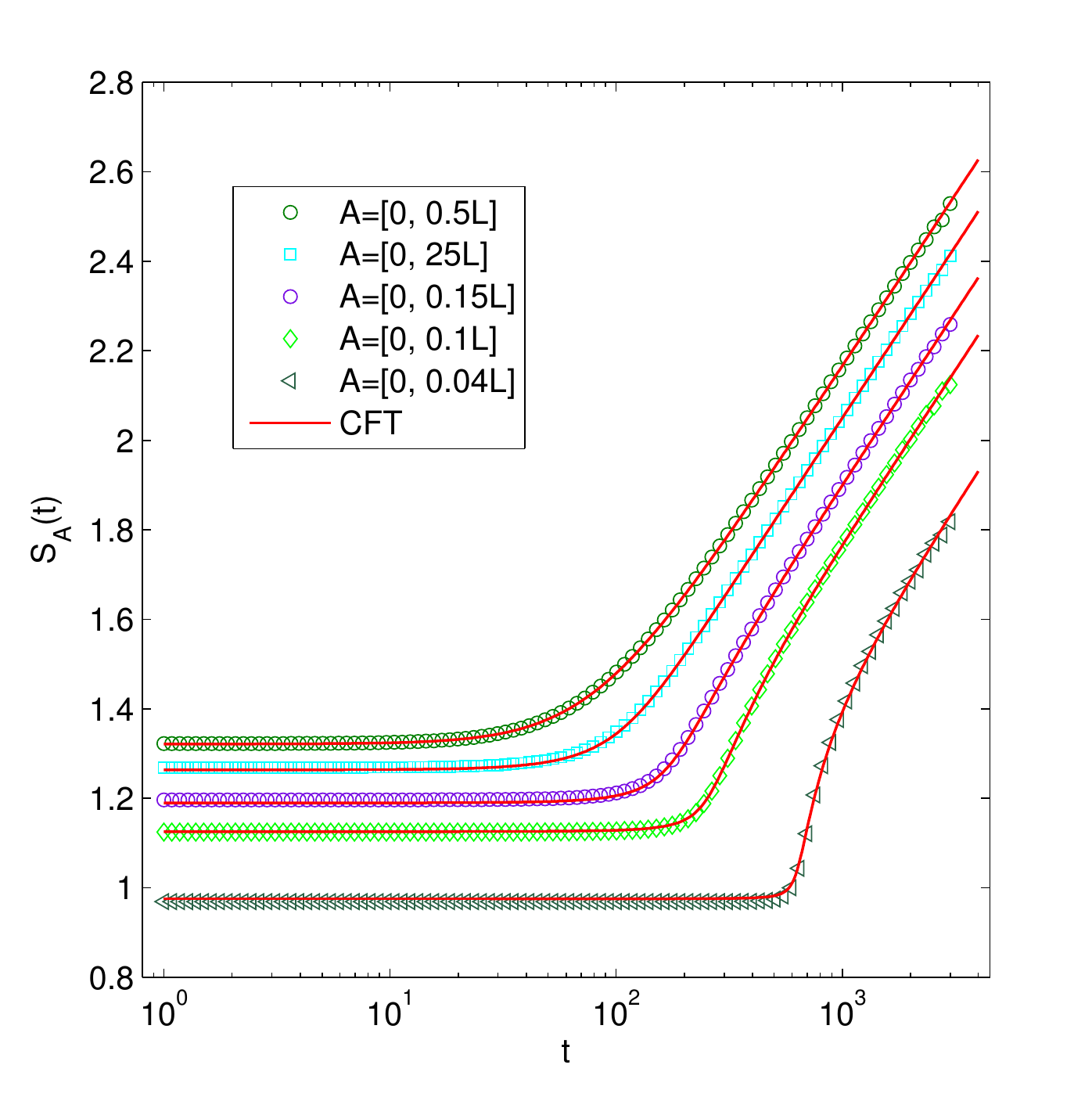}
\caption{
Comparison between numerical results and CFT calculation for
entanglement entropy evolution $S_A(t)$ after a quantum quench from a uniform CFT to a
sine-square deformed CFT. Here we choose $L=500$.}
\label{SSD_quench_fig}
\end{figure}

(i) $t=0$

This corresponds to the ground state of a CFT with open boundary condition. $S_A(t)$
 in Eq.\eqref{SAt_SSD} can be simplified as
\be\label{SAt=0}
S_A(t=0)=\frac{c}{6}\log
\left[
\frac{L}{\pi a}\sin\left(\frac{\pi l}{L}\right)
\right].
\ee
This is the well known result for a finite interval of length $l$ at the end of a CFT living on $[0,L]$.\cite{CC2004}

(ii) $l=L/2$, $t>0$

Now the subsystem is the left half (or right half) of the total system. $S_A(t)$ has a very simple expression
as follows
\be\label{HalfL}
S_A(t)=\frac{c}{6}\log \left[\frac{1}{a^2}\Big(
t^2+\frac{L^2}{4\pi^2}\Big)
\right]-\frac{c}{6}\log \left(
\frac{L}{4\pi a}
\right).
\ee
One can find there is a crossover around $t^{\ast}=L/2\pi$ (see also Fig.\ref{SSD_quench_fig}).
For $t\ll t^{\ast}$, $S_A(t)=\frac{c}{6}\log(L/\pi a)$ is independent of time;
while for $t\gg t^{\ast}$, $S_A(t)\simeq \frac{c}{3}\log t$, which is universal and
independent of $l$ and $L$, as can be observed in Fig.\ref{SSD_quench_fig}.

(iii) $l\ll L$, $t>0$

As shown in in Fig.\ref{SSD_quench_fig},
for arbitrary $l\in (0,L)$, there is a crossover time $t^{\ast}$, so that for $t\ll t^{\ast}$ one has
$S_A(t)\simeq S_A(t=0)$, and for $t\gg t^{\ast}$ one has $S_A(t)\simeq \frac{c}{3}\log t$.
In particular, for $l\ll L$, one can find a simple expression of $t^{\ast}$ as follows:
\be\label{t_star}
\boxed{
t^{\ast}=\frac{L^2}{2\pi^2 l}}\,,\quad l\ll L.
\ee
When the length $L$ of total system is fixed, one can find that $t^{\ast}\propto l^{-1}$.
This explains why there is a wider plateau for smaller $l$ in Fig.\ref{SSD_quench_fig}.
In other words, the smaller $l$ is, the longer time $S_A(t)$ stays in its initial value $S_A(t=0)$.
This may be intuitively understood as follows. Since the Hamiltonian density is sine-square
deformed, the local group velocity of quasi-particles also varies in position.
The group velocity is smaller near the boundary, and larger near the
center of the system. If the entanglement cut is close to the boundary (this corresponds to
$l\ll L$ or $L-l\ll L$), it takes the quasi-particles longer time to reach (or escape) subsystem $A$.
Therefore, $S_A(t)$ will stay at its initial value for a longer time.

For general $l\in (0,L)$, the crossover time $t^{\ast}$ is determined by
\be\label{t_star_general}
t^{\ast}=\frac{L}{2\pi}\cdot \sin^{-1}\Big(\frac{\pi l}{L}\Big)\cdot
\sqrt{
\text{max}\left(
\Big|\cos\frac{2\pi l}{L}\Big|, \,\Big|\sin\frac{2\pi l}{L}\Big|
\right)
}.
\ee
We emphasize that for $t\gg t^{\ast}$, the entanglement entropy grows as $S_A(t)\simeq \frac{c}{3}\log t$
all the way, with no revival. This is the feature of an infinite system.
On the other hand, in the works by Ishibashi and Tada,\cite{Tada1504,Tada1602}, it was found that
the energy spectrum of a CFT with SSD is continuous. Recall that a uniform CFT on a finite space of length
$L$ has energy spacing $\sim 1/L$, which is discrete for a finite $L$. From this point of view, a CFT with SSD
seems to have an infinite length limit.
Here, we studied this effect from the time evolution of entanglement entropy after a quantum quench.

As a remark, it is noted that in the CFT calculation the entanglement entropy grows as
$S_A(t)\simeq \frac{c}{3}\log t$ with no upper bound in the long time limit.
Apparently, this is not the case for a lattice model, since there is a finite
number of degrees of freedom in a subsystem of finite length and the 
energy spectrum is of finite width.
In a lattice model, the entanglement entropy will finally saturate.
One can refer to Ref.\onlinecite{Wen2018Floquet} for more related discussions.

\subsection{Physical interpretation of $t^{\ast}$}
\label{Sec: t*}

To further understand the physical meaning of $t^{\ast}$ in Eq.\eqref{t_star}, again, it is helpful to
consider the quasi-particle picture.
Compared to the cases of
global quench,\cite{calabrese2005evolution,CC_Global,CC2016} local quench,\cite{CC_Local,Nozaki2014,He2014,Stephan2011} and inhomogeneous quantum quench, \cite{Cardy2008,Viti1507,Dubai1606,Eisler2017,BertiniPRL2016,Dubai2017b,BertiniPRL2018,wen2018entanglement}
there is a fundamental difference here. In our case,
since the initial state is the ground state of a uniform CFT, which is long-range entangled,
there is not an intuitive picture on how the entangled pairs of quasi-particles are distributed
in the initial state.
\footnote{It is noted that in the global quench, \cite{calabrese2005evolution,CC_Global,CC2016} since the initial state
is short-range entangled, the entangled-pairs in the initial state are localized in space.
In the local quench in Ref.\onlinecite{CC_Local},
the entangled paris are emitted from the region where two CFTs are connected.
In both cases, we know clearly the distribution of entangled pairs in the initial state.
}

To discuss the physical meaning of $t^{\ast}$, we assume that the quasi-particles are
emitted from the main bulk of the system,
and then we check the time scale that these quasi-particles
propagate into the region $(0,l)$ with $l\ll L$.
This assumption is quite reasonable because the Hamiltonian density
$2\sin^2\left(\frac{\pi x}{L}\right)h(x)$ is more uniform near the two ends
of the SSD system, and looks almost the same as the uniform Hamiltonian density $h(x)$
up to a global factor.
Then in the quantum quench by evolving the ground state of $H_0$ with $H_{\text{SSD}}$,
the quasi-particles are mainly emitted from the bulk of the system.

Now we consider the quasiparticles emitted from $\xi$, with $\xi\sim \mathcal{O}(L)$.
These quasi-particles will contribute to the entanglement entropy of $A=(0,l)$ after a time
\be
t_q=\int_l^{\xi}\frac{dx}{v(x)}.
\ee
Here $v(x)=2\sin^2\left(\frac{\pi x}{L}\right)$ is the group velocity of quasi-particles at $x$.
It is straightforward to obtain
\be
t_q=\frac{L}{2\pi}\left(
\frac{1}{\tan \frac{\pi l}{L}}-\frac{1}{\tan \frac{\pi \xi}{L}}
\right).
\ee
Recall that $l\ll \xi$ and $l\ll L$, then $t_q$ can be further simplified as
\be
t_q\simeq \frac{L^2}{2\pi^2 l},
\ee
which is nothing but $t^{\ast}$ in Eq.\eqref{t_star}.
That is, $t^{\ast}$ actually characterizes the light-cone of quasiparticles that enter the subsystem $A=(0,l)$.
This explains why for $t\ll t^{\ast}$, the entanglement entropy $S_A(t)$ of $A=(0,l)$ does not increase,
while for $t\sim t^{\ast}$, the entanglement entropy starts to grow in time.

For a generic $l$ which is of order $\mathcal{O}(L)$,
to have a quasi-particle interpretation of $t^{\ast}$ in Eq.\eqref{t_star_general},
one needs to know more concrete information about the distribution
of entangled pairs of quasi-particles in the initial state,
which is beyond the scope of our current work.

\section{Entanglement evolution:\\
Quench from uniform to M$\ddot{\text{o}}$bius deformed systems}
\label{QuenchToMobius}

\begin{figure}[tp]
\begin{center}
\includegraphics[width=3.3in]{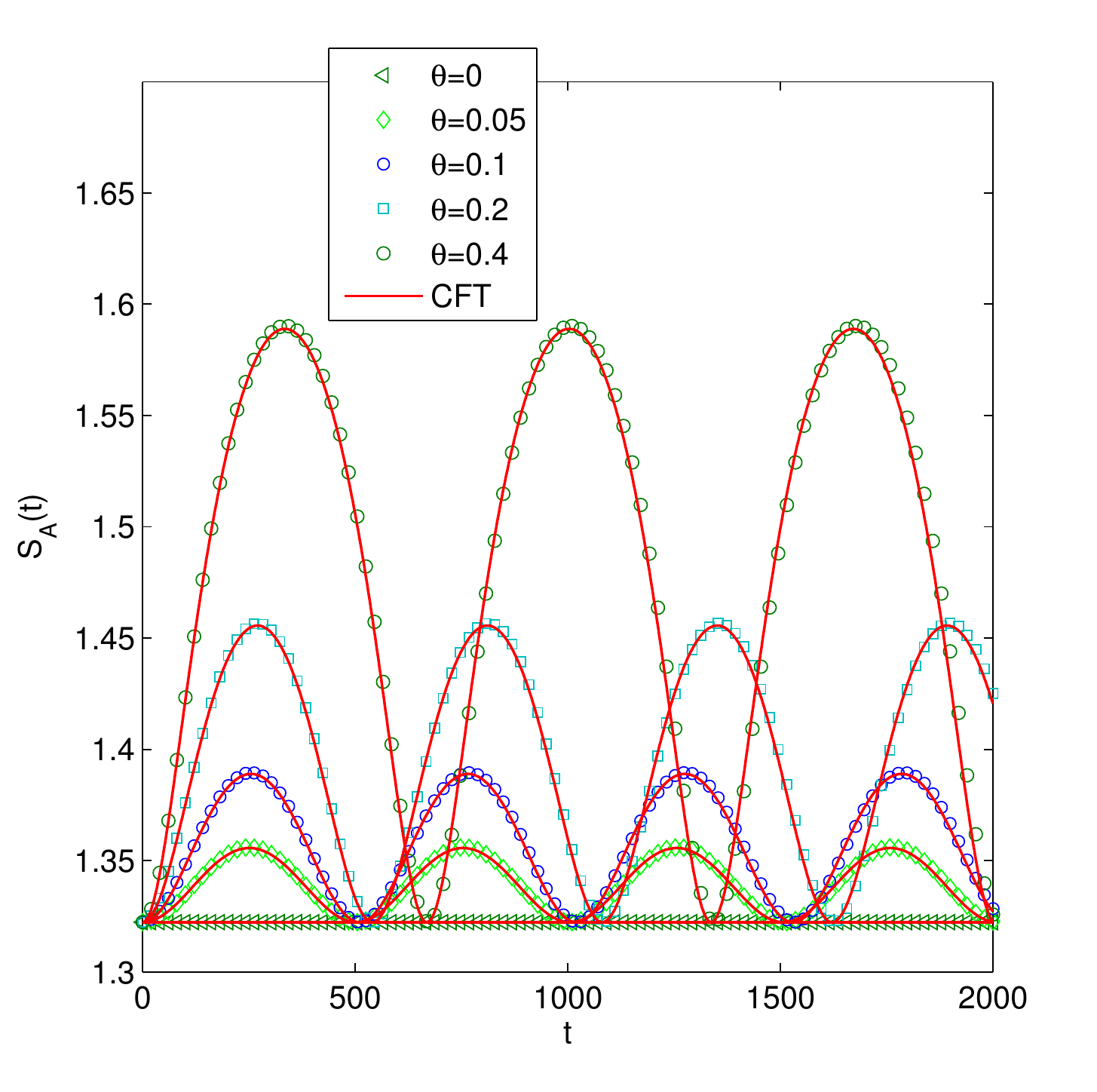}
\end{center}
\caption{
Comparison between numerical results and CFT calculation for
entanglement entropy evolution after a quantum quench from a uniform CFT to a
M$\ddot{o}$bius deformed CFT.
It is found that the period of oscillations is $L\cosh(2\theta)$ and the amplitude of oscillation is $2\theta/3$.
Here we choose $l=L/2$ with $L=500$.
}
\label{Mobius_plot}
\end{figure}

In the previous section, we have studied the quantum quench from a uniform system to a SSD system.
It is natural to ask the following: what happens for the M$\ddot{\text{o}}$bius deformation?
It is interesting to see how the M$\ddot{\text{o}}$bius deformation interpolates between the
uniform and SSD cases.

To be concrete, we prepare the initial state as the ground state of a uniform CFT on a finite space $[0,L]$.
Starting from $t=0$, we have the initial state evolve according to the M$\ddot{\text{o}}$bius Hamiltonian
in Eq.\eqref{MobiusHamiltonian}, and we study the time evolution of entanglement entropy.
The procedures are almost the same as those in Sec.\ref{QuenchToSSD}, except that now we do not take the
limit $\theta\to \infty$. After some straightforward algebra, one can find the time evolution of entanglement entropy
for subsystem $A=[0,l]$ as
\be\label{SAtMobius01}
S_A(t)=\frac{c}{12}\log\left\{
\frac{L^2}{2\pi^2a^2}
\left[
f(t)^2+f(t)\cdot h(t)
\right]
\right\},
\ee
where, as before, we have neglected the $\mathcal{O}(1)$ contribution from the conformal boundary condition.
$f(t)$ and $h(t)$ have the expressions:
\be
f(t)=\sqrt{h(t)^2+\sin^2\frac{2\pi l}{L}},
\ee
and
\be
\begin{split}
h(t)=&-\left(
\sin^2\frac{\pi t}{L_{\text{eff}}}\cdot \cosh(4\theta)+\cos^2\frac{\pi t}{L_{\text{eff}}}
\right)\cdot \cos\frac{2\pi l}{L}\\
&+\sin^2\frac{\pi t}{L_{\text{eff}}}\cdot \sinh(4\theta),
\end{split}
\ee
where
\be\label{Leff}
L_{\text{eff}}=L\cosh(2\theta).
\ee
This effective length can be alternatively obtained by considering a CFT in curved spacetime (see Appendix \ref{Appendix_curve}).
As a self-consistent check, one can find that $S_A(t)$ reduces to Eq.\eqref{SAt=0} for $t=0$,
and reduces to Eq.\eqref{SAt_SSD} for $\theta\to \infty$, as expected.

The comparison between CFT results and the numerical calculation is shown in Fig.\ref{Mobius_plot}, and
the agreement is excellent.
For $\theta=0$, since the M$\ddot{\text{o}}$bius Hamiltonian is the same as the uniform case, then
there is essentially no quench (see Fig.\ref{Mobius_plot}).
For $\theta>0$, it is interesting that oscillations appear in $S_A(t)$.
Based on $S_A(t)$ in Eqs.\eqref{SAtMobius01}$\sim$\eqref{Leff}, one can find the period of
oscillations is $L_{\text{eff}}=L\cosh(2\theta)$.
For $\theta\to 0$, the oscillation period is $L$, which is as expected for a uniform CFT.
On the other hand, in the SSD limit $\theta\to \infty$,
the oscillation period becomes $L\cosh(2\theta)\to \infty$.
This agrees with the fact that there is no revival in $S_A(t)$ for a CFT with SSD.

To see the features of $S_A(t)$ in Eq.\eqref{SAtMobius01} more clearly, let us focus on the
case $l=L/2$. Then $S_A(t)$ can be simplified as
\be\label{MobiusLAmiddle}
S_A(t)=\frac{c}{6}\log
\left\{
\frac{L}{\pi a}\left[\sin^2\left(\frac{\pi t}{L_{\text{eff}}}\right)e^{4\theta}+\cos^2\left(\frac{\pi t}{L_{\text{eff}}}\right)\right]
\right\}.
\ee
The period of oscillations with $T=L_{\text{eff}}=L\cosh(2\theta)$ can be explicitly seen in this
expression. In addition, one can find that the amplitude of oscillations grows as we increase $\theta$.
For $l=L/2$, the amplitude of oscillations has a very simple expression
\be
S_A(t=\frac{L_{\text{eff}}}{2})-S_A(t=0)=\frac{2c}{3}\theta,
\ee
which grows with $\theta$ linearly, as can be observed in Fig.\ref{Mobius_plot}.

Furthermore, it is straightforward to check that as $\theta\to \infty$, the entanglement entropy evolution in
Eq.\eqref{SAtMobius01} will reduce to the SSD case in Eq.\eqref{SAt_SSD}, as expected.

As a short summary, by studying the quench dynamics in a CFT with M$\ddot{\text{o}}$bius deformation,
one can find that the effective length of the system becomes $L_{\text{eff}}=L\cosh(2\theta)$,
which interpolates between the uniform and SSD systems, as we tune $\theta$ from $0$ to $\infty$.

\textit{Remark:}
It is noted that the effective length $L_{\text{eff}}$ can also be understood based on the quasiparticle picture.
For the M$\ddot{\text{o}}$bius deformation, the group velocity of quasi-particles at $x$ is
$v(x)=1-\tanh(2\theta)\cos\frac{2\pi x}{L}$.
Since the system is symmetric about $x=L/2$,
for an entangled pair of quasiparticles emitted from $x$, they will meet again at $L-x$ at time
\be
t_q=\int_0^L\frac{dx}{v(x)}=L\cosh(2\theta).
\ee
Similarly, the entangled-pair of quasiparticles emitted from $L-x$ will meet again for the first time
at $x$ with time $t_q=L\cosh(2\theta)$.
This explains why we observe a revival of $S_A(t)$ with a time period $L\cosh(2\theta)$.

\section{Concluding remarks}
\label{Conclusion}

Let us first summarize our main results, and then make some comments.

We studied analytically the quench dynamics of a sine-square deformed CFT,
which was proposed to have continuous energy spectrum and infinite length limit.
By quenching from a uniform CFT to a sine-square deformed CFT on $[0,L]$, it was found that
the entanglement entropy $S_A(t)$ for subsystem $A=[0,l]$ evolves in time in a universal way.
There exists a crossover time $t^{\ast}$. For $t\ll t^{\ast}$, $S_A(t)$
does not evolve in time; for $t\gg t^{\ast}$, $S_A(t)$ grows in time all the way as
$S_A(t)\simeq \frac{c}{3}\log t$, with no revival. This feature indicates that the CFT with SSD
effectively has an infinite length limit, consistent with previous analysis
on the energy spectrum.
In addition, we studied analytically the quench dynamics of a
M$\ddot{\text{o}}$bius deformed CFT.
Aside from some interesting features, it was found that a length
scale $L_{\text{eff}}=L\cosh(2\theta)$ appears in the time evolution
of the entanglement entropy, which interpolates between the uniform
and SSD systems. We hope our work can stimulate more interest
in the non-equilibrium dynamics in sine-square deformed CFTs and other
related non-uniform CFTs.

As will be studied in Ref.\onlinecite{Wen2018Floquet}, the setup in this work
provides a building block for studying the Floquet dynamics in a conformal field theory.
That is, one can drive a CFT with Hamiltonians $H_0$ and $H_{\text{SSD}}$
[see Eqs.\eqref{H00} and \eqref{SSD}] periodically, and see if the system can
be heated or not.
Compared to the previous work on boundary-driven CFT, \cite{FloquetBCFT}
now we have a bulk-driven Floquet CFT which can be analytically solved.\cite{Wen2018Floquet}

In addition, careful readers may have noticed that when evaluating the correlation function in Eq.\eqref{Tn_Eu},
we did not introduce any UV regularization, in contrast to other setups such as Refs.\onlinecite{calabrese2005evolution,CC_Global,CC2016,CC_Local,Nozaki2014,He2014,Stephan2011}.
The reason is as follows. The effect of time evolution operator $U(t)=e^{-iH_{\text{SSD}}t}$
is simply to evolve the primary operator $\mathcal{O}(z,\bar{z})$ to $\mathcal{O}(z_{\text{new}},\bar{z}_{\text{new}})$.
The rest of the calculation is essentially evaluating the correlation function within the \textit{ ground state} of $H_0$.
This is in contrast to other setups where the evaluation of correlation functions cannot be reduced
to the calculation within the ground state of $H_0$.

There are many open questions and we would like to mention some of them here:

-- In our work, we quench a uniform CFT to a non-uniform CFT.
There are many other  interesting setups for quantum quenches, which may be used to
study the property of sine-square deformed CFTs, such
as the global and local quenches as introduced in Refs.\onlinecite{CC_Global,CC_Local}.
Technically, it will be more involved to study these setups because the CFT with SSD is defined
on a finite space.
There are more boundaries introduced by the global/local quenches,
and one needs complicate conformal mappings to study these quenches.
We also want to point out that there are other interesting entanglement measures that may be
helpful to detect how the entanglement is generated (or propagates)
in a CFT with SSD. See, \textit{e.g.}, Fig.10 in Ref.\onlinecite{Wen2016EN}, on how to
use entanglement negativity to detect the distribution of EPR pairs in a CFT after a quantum quench.

-- It is also interesting to study other kinds of deformations, such as
a $\sin^n$ deformation. It is expected that the Virasoro generators
$L_n$ and $L_{-n}$ with $n>1$ will also appear in the Hamiltonian.
The feature of energy spectrum for CFTs with such deformations was even
not well understood. It is interesting to see if such deformations
can be analytically studied within the CFT approach.

-- Recently, measuring the time evolution of (Renyi) entanglement entropy after a quantum quench
 was realized in cold atom experiments,\cite{Kaufman794}
where the (1+1)-d quantum system is quenched from a Mott insulator phase to a superfluid phase.
Here, the setup in our work applies to arbitrary (1+1)-d quantum critical systems that can be described by
a CFT. We expect that our setup may be realized in cold-atom experiments by tuning the tunneling strength
between neighboring sites through the optical lattice depth. \cite{bakr_2009}
It is noted that the tunneling strength corresponds to the hopping strength in the lattice model (see Appendix A).
By tuning the tunneling strength in space in a sine-square deformed way, 
one may realize SSD as well as its quech dynamics in experiments.

\section{Acknowledgement}

We thank Shinsei Ryu for helpful discussions on various properties of SSD,
and thank Chenjie Wang for discussions on CFT in curved spacetime which stimulates our interest
in studying this problem, and introducing their recent work [\onlinecite{Gu1801}] to us.
We also thank Erik Tonni for discussions on entanglement entropy of
2d CFT in curved spacetime, and thank Wenchao Xu for discussions on
the possible realization of our setup in cold atom experiments.
XW is supported by the Gordon and Betty Moore Foundation's EPiQS initiative
through Grant No. GBMF4303 at MIT.
JQW is supported by Massachusetts Institute of Technology and  the Simons foundation it from qubit collaboration.

\appendix

\section{Lattice model}
\label{Appendix_Lattice}

To confirm our field theory result in the main text,
we calculate the entanglement entropy evolution based on a free fermion lattice model.
We prepare the initial state as the ground state $|G\rangle$ of a uniform free fermion chain with half-filling:
\be
H_0=\sum_{i=1}^{L-1} h c_i^{\dag}c_{i+1}+h.c.,
\ee
where $h$ is the hopping strength, and we choose $h=1/2$ throughout the calculation.
The length of the chain is $L$ and open boundary condition is imposed.
$c_i$ ($c_i^{\dag}$) are fermionic operators, which satisfy the
anticommutation relations $\{c_i,c_j\}=\{c_i^{\dag},c_j^{\dag}\}=0$, and
$\{c_i,c_j^{\dag}\}=\delta_{ij}$.

Then at time $t=0$, we have the initial state $|G\rangle$ evolve
according to the new hamiltonian, which is non-uniform in space:
\be
H_1=\sum_{i=1}^{L-1}\left[
1-\tanh(2\theta)\cdot \cos\left(\frac{2\pi (i+1/2)}{L}\right)
\right]c_i^{\dag}c_{i+1}+h.c.
\ee
Note that for $\theta=0$, $H_1$ corresponds to the uniform Hamiltonian, and for
$\theta\to \infty$, $H_1$ corresponds to the Hamiltonian with SSD.
Then one can calculate the two-point correlation function
$\langle \psi(t)|c_i^{\dag}c_j|\psi(t)\rangle$ in subsystem $A$,
with $|\psi(t)\rangle=e^{-iH_1t}|G\rangle$.
Based on the two-point correlation functions, one can calculate the entanglement
entropy $S_A(t)$ following Peschel's method.\cite{Peschel2003}

We compare our numerical calculation with the CFT results for cases
with $\theta\to\infty$ and finite $\theta$, respectively.
The only fitting parameter we choose is the global shift (which is a constant) in the ground-state
entanglement entroy $S_A(t=0)$,
arising from the cut-off and boundary conditions. (It is noted that for a free fermion model, this
constant term in $S_A(t=0)$ has been explicitly evaluted in Ref.\onlinecite{fagotti2011universal}.)
The agreement is excellent, as shown in Figs.\ref{SSD_quench_fig} and \ref{Mobius_plot}.

\section{CFT in curved space-time}
\label{Appendix_curve}

In this section, we explain that the sine-square deformed
Hamiltonian or M$\ddot{\text{o}}$bius Hamiltonian can be regarded as
a CFT in curved space-time.

The CFT in curved space is invariant under coordinate transformation and Weyl transformation. For example, we consider a multi-point correlation function
\be \langle O_1(x_1)O_2(x_2)...\rangle_{ds^2_1}, \ee
in the space with metric
\be ds^2_1=g_{\mu\nu}(x)dx^{\mu}dx^{\nu}. \ee
The correlation function is invariant under the coordinate transformation
\be\label{tr1}
\begin{split}
&O_i(x_i)\rightarrow O_i(y_i)\mid_{y_i=f(x_i)}  \\
& ds_1^2\rightarrow ds_2^2=g_{\mu\nu}(x)\frac{\partial x^{\rho}}{\partial y^{\mu}}\frac{\partial x^{\sigma}}{\partial y^{\nu}}\mid_{x=f^{-1}(y)},
\end{split}
\ee
and the Weyl transformation
\be\label{tr2}
\begin{split}
&O_i(x_i)\rightarrow e^{-\Delta_i \sigma(x_i)}O_i(x_i) \\
& ds_1^2\rightarrow ds_3^2=e^{2\sigma(x)}ds_1^2,
\end{split}
\ee
where $\Delta_i=h_i+\bar{h}_i$. More explicitly, the three correlation functions
\be
\begin{split}
& \langle O_1(x_1)O_2(x_2)...\rangle_{ds^2_1} \\
=& \langle O_1(y_1)O_2(y_2)...\rangle_{ds^2_2}\mid_{y_i=f(x_i)} \\
=& e^{-\sum_i \Delta_i\sigma(x_i)}\langle O_1(x_1)O_2(x_2)...\rangle_{ds^2_3} \\
\end{split}
\ee
are equal to each other. With these results, we can rewrite the theory with
M$\ddot{\text{o}}$bius Hamiltonian as a CFT in curved space.

As discussed in section \ref{QuenchToSSD}, the operators in coordinate $w$ and coordinate $\tilde{z}$ are related by
\be\label{operator}
\begin{split}
&\phi^{(w)}(t_E,x)=e^{Ht_E}\phi^{(w)}(0,x) e^{-Ht_E}\\
=& (\frac{2\pi}{L})^{2h}
\frac{e^{\frac{4\pi h}{L\cosh2\theta}t_E}}{(\cosh 2\theta-\sinh2\theta\cos\frac{2\pi x}{L})^{2h}}\phi^{(\tilde{z})}(\tilde{z},\bar{\tilde{z}}),
\end{split}
\ee
where
\be\label{coor}
\begin{split}
& \tilde{z}=-e^{\frac{2\pi t_E}{L\cosh 2\theta}}
\frac{\cosh\theta e^{\frac{2\pi i}{L}x}-\sinh\theta}{\sinh\theta e^{\frac{2\pi i}{L}x}-\cosh\theta}, \\
& \bar{\tilde{z}}=-e^{\frac{2\pi t_E}{L\cosh 2\theta}}
\frac{\cosh\theta e^{\frac{-2\pi i}{L}x}-\sinh\theta}{\sinh\theta e^{-\frac{2\pi i}{L}x}-\cosh\theta}.
\end{split}
\ee

We will show that the M$\ddot{\text{o}}$bius Hamiltonian in $w$ coordinate can be regarded as CFT in the space with metric
\be ds_{(w)}^2=\frac{dt_E^2}{\cosh^2 2\theta}(\cosh 2\theta-\sinh 2\theta\cos\frac{2\pi x}{L})^2+dx^2. \ee
Note that the metric in $\tilde{z}$ is
\be ds_{\tilde{z}}^2=d\tilde{z}d\bar{\tilde{z}}. \ee
The metric in $\tilde{z}$ and $w$ can be transformed to each other by a coordinate transformation (\ref{coor}) and a Weyl transformation
\be ds_{(\tilde{z})}^2=e^{2\sigma}ds_{(w)}^2, \ee
where
\be e^{2\sigma}=(\frac{2\pi}{L})^2\frac{e^{\frac{2\pi t_E}{L\cosh2\theta}}}{(\cosh2\theta-\sinh2\theta\cos\frac{2\pi x}{L})^2}. \ee
With (\ref{tr1}) and (\ref{tr2}), we get the same relation (\ref{operator}) from CFT in curved space.
Furthermore, we can calculate the effective length of the system
\be \int_0^{L} \sqrt{\frac{g_{xx}}{g_{tt}}} dx=L\cosh 2\theta, \ee
which is the same as $L_{\text{eff}}$ in Eq.\eqref{Leff}.
As studied in Sec.\ref{Sec: t*},
it is also interesting to check the effective distance between $x_1=l$ and $x_2=\xi$ as follows:
\be
\begin{split}
L_{\text{eff}}(x_1,x_2;\theta\to\infty)&=
 \int_l^{\xi} \sqrt{\frac{g_{xx}}{g_{tt}}} dx\\
&=\frac{L}{2\pi}\left(
\frac{1}{\tan \frac{\pi l}{L}}-\frac{1}{\tan \frac{\pi \xi}{L}}
\right).
\end{split}
\ee
For $l\ll L$ and $l\ll \xi$, one has
\be
L_{\text{eff}}(x_1,x_2;\theta\to\infty)\simeq \frac{L^2}{2\pi^2 l}.
\ee


\bibliography{SSDref}

\end{document}